\documentstyle[12pt]{article}
\def\be {\begin{equation}}
\def\ee {\end{equation}}
\def\ba {\begin{eqnarray}}
\def\ea {\end{eqnarray}}

\def\bi {\begin{itemize}}
\def\ei {\end{itemize}}
\begin{document}
\def\bea{\begin{eqnarray}}
\def\eea{\end{eqnarray}}
\title{\bf {Interacting holographic dark energy in the scalar-Gauss-Bonnet gravity}}
 \author{M.R. Setare  \footnote{E-mail: rezakord@ipm.ir}
  \\ {Department of Science,  Payame Noor University. Bijar, Iran}}
\date{\small{}}

\maketitle
\begin{abstract}
In this paper we study cosmological application of interacting
holographic dark energy density in the scalar-Gauss-Bonnet
framework. We employ
 the interacting holographic model of dark energy to obtain the equation of state  for the interacting
  holographic energy density
 in spatially flat universe. Our calculation show, taking $\Omega_{\Lambda}=0.73$ for
the present time, it is possible to have $w_{\rm \Lambda}^{eff}$
crossing $-1$. This implies that one can generate phantom-like
equation of state from the interacting holographic dark energy model
in flat universe in the scalar-Gauss-Bonnet cosmology framework.
Then we reconstruct the potential of the scalar field.
 \end{abstract}
% \begin{document}

\newpage
% \vspace*{10mm}

\section{Introduction}
Nowadays it is strongly believed that the universe is experiencing
an accelerated expansion. Recent observations from type Ia
supernovae \cite{SN} in associated with Large Scale Structure
\cite{LSS} and Cosmic Microwave Background anisotropies \cite{CMB}
have provided main evidence for this cosmic acceleration. In order
to explain why the cosmic acceleration happens, many theories have
been proposed. It is the most accepted idea that a mysterious
dominant component, dark energy, with negative pressure, leads to
this cosmic acceleration, though its nature and cosmological origin
still remain enigmatic at present. An alternative proposal for dark
energy is the dynamical dark energy scenario. The cosmological
constant puzzles may be better interpreted by assuming that the
vacuum energy is canceled to exactly zero by some unknown mechanism
and introducing a dark energy component with a dynamically variable
equation of state. The dynamical dark energy proposal is often
realized by some scalar field mechanism which suggests that the
energy form with negative pressure is provided by a scalar field
evolving down a proper potential.\\
In recent years, many string theorists have devoted to understand
and shed light on the cosmological constant or dark energy within
the string framework. The famous Kachru-Kallosh-Linde-Trivedi (KKLT)
model \cite{kklt} is a typical example, which tries to construct
metastable de Sitter vacua in the light of type IIB string theory.
Furthermore, string landscape idea \cite{landscape} has been
proposed for shedding light on the cosmological constant problem
based upon the anthropic principle and multiverse speculation.
Although we are lacking a quantum gravity theory today, we still can
make some attempts to probe the nature of dark energy according to
some principles of quantum gravity. The holographic dark energy
model is just an appropriate example, which is constructed in the
light of the holographic principle of quantum gravity theory. That
is to say, the holographic dark energy model possesses some
significant features of an underlying theory of dark energy.
Currently, an interesting attempt for probing the nature of dark
energy within the framework of quantum gravity is the so-called
``holographic dark energy'' proposal
\cite{Cohen:1998zx,Horava:2000tb,Hsu:2004ri,Li:2004rb}. It is well
known that the holographic principle is an important result of the
recent researches for exploring the quantum gravity (or string
theory) \cite{holoprin}. Such a holographic dark energy looks
reasonable, since it may provide simultaneously natural solutions to
both dark energy problems as demonstrated in Ref.\cite{Li:2004rb}.
The holographic dark energy model has been tested and constrained by
various astronomical observations \cite{obs3}. Furthermore, the
holographic dark energy model has been extended to include the
spatial curvature contribution, i.e. the holographic dark energy
model in non-flat space \cite{nonflat}. It is known that the
coincidence or, ``why now" problem is easily solved in some models
of holographic dark energy based on this fundamental assumption that
matter and holographic dark energy do not conserve separately, but
the matter energy density decays into the holographic energy density
\cite{interac}. In fact a suitable evolution of the Universe is
obtained when, in addition to the holographic dark energy, an
interaction (decay of
dark energy to matter) is assumed.\\
 Because the holographic energy density belongs to a
dynamical cosmological constant, we need a dynamical frame to
accommodate it instead of general relativity. Therefore it is
worthwhile to investigate the holographic energy density in the
framework of the Brans-Dicke theory \cite{{gong},{mu},
{tor},{set1}}. Einstein's theory of gravity may not describe gravity
at very high energy. The simplest alternative to general relativity
is Brans-Dicke scalar-tensor theory \cite{bd}. But among the most
popular modified gravities which may successfully describe the
cosmic speed-up is $F(R)$ gravity. Very simple versions of such
theory like $1/R$ \cite{1} and $1/R + R^2$ \cite{2} may lead to the
effective quintessence/phantom late-time universe. Another theory
proposed as gravitational dark energy is scalar-Gauss-Bonnet gravity
\cite{3}
which is closely related with low-energy string effective action.\\
In present paper, using the interacting holographic model of dark
energy in spatially flat universe, we obtain equation of state for
holographic dark energy density for a scalar-Gauss-Bonnet universe
enveloped by  $R_h$ as the system's IR cut-off. The current
available observational data imply that the holographic vacuum
energy behaves as phantom-type dark energy, i.e. the
equation-of-state of dark energy crosses the cosmological-constant
boundary $w=-1$ during the evolution history. We show this phantomic
description of the holographic dark energy in flat universe with
$0.21\leq c\leq 2.1$, then we reconstruct the potential of the
phantom scalar field.
\section{Holographic dark energy in Gauss-Bonnet  Framework}

 The action containing Gauss-Bonnet interaction is given
 by
\begin{equation}
S+S_m=\int d^{4}x\,\sqrt{g}\,\left[
\frac{1}{2k^2}R-\frac{\gamma}{2}\partial_{\mu}\phi\partial^{\mu}\phi-V(\phi)+f(\phi)G\right]+S_m
. \label{action*}
\end{equation}
where $\gamma=\pm1$ $\phi$ is a scalar field and
$G=R^2-4R_{\mu\nu}R^{\mu\nu}+R_{\mu\nu\rho\sigma}R^{\mu\nu\rho\sigma}$
is the Gauss-Bonnet term which is appearing in the action with a
coupling parameter $f(\phi)$, also $S_m$ is the action of pressureless Cold Dark Matter (CDM).\\
For the spatially flat Robertson-Walker universe
\begin{equation}\label{met}
ds^{2}=-dt^{2}+a(t)^{2}(dr^{2}+r^{2}d\Omega^{2}).
\end{equation}
The field equations are (here we assume $\frac{1}{k^2}=M_{P}^{2}=1$)
\begin{equation}\label{1}
\rho+\rho_{m}=\frac{\gamma}{2}\dot{\phi}^{2}+V(\phi)-24f'(\phi)\dot{\phi}H^3
\end{equation}
\begin{equation}\label{2}
\frac{\gamma}{2}\dot{\phi}^{2}-V(\phi)+16f'(\phi)\dot{\phi}H\frac{\ddot{a}}{a}+8(f'(\phi)\ddot{\phi}
+f''(\phi)\dot{\phi}^{2})H^2=P
\end{equation}
where, $p$ and $\rho$ are the pressure and energy density due to the
scalar field and the Gauss-Bonnet interaction, and $\rho_{m}$ is the
energy density of CDM. In addition the equation of motion for scalar
field is as
\begin{equation}\label{3}
\gamma(\ddot{\phi}+3H\dot{\phi}+\frac{V'(\phi)}{\gamma})=24f'(\phi)H^2\frac{\ddot{a}}{a}
\end{equation}
We allow for an arbitrary coupling between CDM and the scalar filed
$\phi$. We assume that the filed $\phi$ is coupled to CDM with a
coupling given by $Q=\Gamma \rho_{\Lambda}$. his is a decaying of
the dark energy component into CDM with the decay rate $\Gamma$. The
continuity equations for dark energy and CDM are
\begin{eqnarray}
\label{2eq1}&& \dot{\rho}+3H(1+w)\rho =-Q, \\
\label{2eq2}&& \dot{\rho}_{\rm m}+3H\rho_{\rm m}=Q.
\end{eqnarray}
Taking a ratio of two energy densities as $r=\rho_{\rm m}/\rho_{\rm
\Lambda}$, the above equations lead to
\begin{equation}
\label{2eq3} \dot{r}=3Hr\Big[w_{\rm \Lambda}+
\frac{1+r}{r}\frac{\Gamma}{3H}\Big]
\end{equation}
 Following Ref.\cite{Kim:2005at},
if we define
\begin{eqnarray}\label{eff}
w ^{\rm eff}=w+{{\Gamma}\over {3H}}\;, \qquad w_m ^{\rm
eff}=-{1\over r}{{\Gamma}\over {3H}}\;.
\end{eqnarray}
Then, the continuity equations can be written in their standard form
\begin{equation}
\dot{\rho} + 3H(1+w^{\rm eff})\rho = 0\;,\label{definew1}
\end{equation}
\begin{equation}
\dot{\rho}_m + 3H(1+w_m^{\rm eff})\rho_m = 0\; \label{definew2}
\end{equation}
 Now we suggest a
correspondence between the holographic dark energy scenario and the
above Gauss-Bonnet dark energy model. The holographic energy density
$\rho_{\Lambda}$ is chosen to be \be
\rho_{\Lambda}=\frac{3c^2}{R_{h}^2} \label{holo}\ee where $c$ is a
constant, and $R_h$ is the future event horizon given by \be
  R_h= a\int_t^\infty \frac{dt}{a}=a\int_a^\infty\frac{da}{Ha^2}
 \ee
 The critical energy density, $\rho_{cr}$, is given by following relation
\begin{eqnarray} \label{ro}
\rho_{cr}=3H^2
\end{eqnarray}
Now we define the dimensionless dark energy as \be
\Omega_{\Lambda}=\frac{\rho_{\Lambda}}{\rho_{cr}}=\frac{c^2}{R_{h}^2H^2}\label{omega}
\ee Using definition $\Omega_\Lambda$ and relation (\ref{ro}),
$\dot{R_{h}}$ gets: \be \label{ldot} \dot{R_{h}} =
R_{h}H-1=\frac{c}{\sqrt{\Omega_\Lambda}}-1,
\end{equation}
By considering  the definition of holographic energy density
$\rho_{\rm \Lambda}$, and using Eq.(\ref{ldot})one can find:
\begin{equation}\label{roeq}
\dot{\rho_{\Lambda}}=\frac{-2}{R_{h}}(\frac{c}{\sqrt{\Omega_\Lambda}}-1)\rho_{\Lambda}
\end{equation}
Substitute this relation into Eq.(\ref{2eq1}) and using definition
$Q=\Gamma \rho_{\Lambda}$, we obtain
\begin{equation}\label{stateq}
w_{\rm \Lambda}=-(\frac{1}{3}+\frac{2\sqrt{\Omega_{\rm
\Lambda}}}{3c}+\frac{\Gamma}{3H}).
\end{equation}
 From Eq.(\ref{eff}), we have the effective
equation of state as
\begin{equation} \label{3eq401}
w_{\rm \Lambda}^{eff}=-\frac{1}{3}-\frac{2\sqrt{\Omega_{\rm
\Lambda}}}{3c}.
\end{equation}
 Here we assume that $\phi=\phi_0a^n$,
where  $\phi_0$, and  $n$ are constant. Using this relation  and
 Eqs.(\ref{1}, \ref{holo}, \ref{omega}), we can  obtain
$\Omega_{\rm \Lambda}$ as \be \label{20}\Omega_{\rm
\Lambda}=\frac{1}{3}[\frac{\gamma}{2}n^2
\phi^{2}+\frac{V(\phi)}{H^2}-24f'n\phi H^{2}] \ee
 Now we consider the case that $f(\phi)$ is given
as \be \label{4} f(\phi)=f_0e^{\frac{2\phi}{\phi_{0}}}\ee where
$f_0$ is a constant. Using Eqs.(\ref{3}, \ref{4}) one can obtain \be
\label{22}
V'(\phi)=\frac{48f(\phi)H^2}{\phi_0}(\dot{H}+H^2)-\gamma[n\phi(\dot{H}+nH^2)+3nH^2\phi]\ee
then we can obtain the scalar potential as \be \label{23}
V(\phi)-V(\phi_{0})=\int_{\phi_{0}}^{\phi}V'(\phi')d\phi'=24H^2(\dot{H}+H^2)(f(\phi)-f_{0}e^{2})-\frac{\gamma
n}{2}[\dot{H}+(3+n)H^2](\phi^{2}-\phi_{0}^{2}) \ee Thus by
considering the above equations, one can see that $w_{\rm
\Lambda}^{eff}$ through dependence to $\Omega_{\rm \Lambda}$ is
depend to the parameters and details of model. $w_{\rm
\Lambda}^{eff}$ also depend to $c$, in the recent fit studies,
different groups gave different values to $c$. A direct fit of the
present available SNe Ia data with this holographic model indicates
that the best fit result is $c=0.21$ \cite{HG}. Recently, by
calculating the average equation of state of the dark energy and the
angular scale of the acoustic oscillation from the BOOMERANG and
WMAP data on the CMB to constrain the holographic dark energy model,
the authors show that the reasonable result is $c\sim 0.7$
\cite{cmb1}. In the other hand, in the study of the constraints on
the dark energy from the holographic connection to the small $l$ CMB
suppression, an opposite result is derived, i.e. it implies the best
fit result is $c=2.1$ \cite{cmb3}. Thus according to these studies
$0.21\leq c\leq 2.1$.  Taking $\Omega_{\Lambda}=0.73$ for the
present time, in the case of $c=0.21$, we obtain $w_{\rm
\Lambda}^{eff}=-3.04$, in the other hand for $c=2.1$, one can
obtain, $w_{\rm \Lambda}^{eff}=-0.6$. Using Eq.(\ref{3eq401}), one
can see that by considering $c\leq \sqrt{\Omega_{\Lambda}}$ we
obtain $w_{\rm \Lambda}^{eff}\leq -1$. Therefore taking
$\Omega_{\Lambda}=0.73$ for the present time, it is possible to have
$w_{\rm \Lambda}^{eff}$ crossing $-1$. This implies that one can
generate phantom-like equation of state from an interacting
holographic dark energy model in flat universe in the
scalar-Gauss-Bonnet cosmology framework.
\section{Conclusions}
In the present paper, by considering the interacting holographic
energy density as a dynamical cosmological constant, we have
obtained the equation of state for the holographic energy density in
the scalar-Gauss-Bonnet framework. We have considered the
holographic dark energy density as Eq.(\ref{holo}), where $c$ is a
positive constant. By choosing $0.21\leq c\leq 2.1$, and taking
$\Omega_{\Lambda}=0.73$ for the present time, we have shown that in
our model it is possible to have $w_{\rm \Lambda}$ crossing $-1$.
Finally we have show this phantomic description of the interacting
holographic dark energy with $c\leq \sqrt{\Omega_{\Lambda}}$ and
reconstruct the potential of the phantom scalar field.


\begin{thebibliography}{99}
\bibitem{SN}
  A.~G.~Riess {\it et al.}  [Supernova Search Team Collaboration],
  %``Observational Evidence from Supernovae for an Accelerating Universe and a
  %Cosmological Constant,''
  Astron.\ J.\  {\bf 116}, 1009 (1998)
  [astro-ph/9805201];\\
  %%CITATION = ASTRO-PH 9805201;%%
%\cite{Perlmutter:1998np}
%\bibitem{Perlmutter:1998np}
  S.~Perlmutter {\it et al.}  [Supernova Cosmology Project Collaboration],
  %``Measurements of Omega and Lambda from 42 High-Redshift Supernovae,''
  Astrophys.\ J.\  {\bf 517}, 565 (1999)
  [astro-ph/9812133];\\
  %%CITATION = ASTRO-PH 9812133;%%
%\cite{Riess:2004nr}
%\cite{Astier:2005qq}
%\bibitem{Astier:2005qq}
  P.~Astier {\it et al.},
   %``The Supernova Legacy Survey: Measurement of Omega_M, Omega_Lambda and w
  %from the First Year Data Set,''
  Astron.\ Astrophys.\  {\bf 447}, 31 (2006)
  [astro-ph/0510447].
  %%CITATION = ASTRO-PH 0510447;%%

%\cite{LSS}
\bibitem{LSS}
  M.~Tegmark {\it et al.}  [SDSS Collaboration],
  %``Cosmological parameters from SDSS and WMAP,''
  Phys.\ Rev.\ D {\bf 69}, 103501 (2004)
  [astro-ph/0310723];\\
  %%CITATION = ASTRO-PH 0310723;%%
  %\cite{Abazajian:2004aj}
%\bibitem{Abazajian:2004aj}
  K.~Abazajian {\it et al.}  [SDSS Collaboration],
  %``The Second Data Release of the Sloan Digital Sky Survey,''
  Astron.\ J.\  {\bf 128}, 502 (2004)
  [astro-ph/0403325];\\
  %%CITATION = ASTRO-PH 0403325;%%
%\cite{Abazajian:2004it}
%\bibitem{Abazajian:2004it}
  K.~Abazajian {\it et al.}  [SDSS Collaboration],
  %``The Third Data Release of the Sloan Digital Sky Survey,''
  Astron.\ J.\  {\bf 129}, 1755 (2005)
  [astro-ph/0410239].
  %%CITATION = ASTRO-PH 0410239;%%

%\cite{CMB}
\bibitem{CMB}
  D.~N.~Spergel {\it et al.}  [WMAP Collaboration],
  %``First Year Wilkinson Microwave Anisotropy Probe (WMAP) Observations:
  %Determination of Cosmological Parameters,''
  Astrophys.\ J.\ Suppl.\  {\bf 148}, 175 (2003)
  [astro-ph/0302209];\\
  %%CITATION = ASTRO-PH 0302209;%%
%\cite{Spergel:2006hy}
%\bibitem{Spergel:2006hy}
  D.~N.~Spergel {\it et al.},
  %``Wilkinson Microwave Anisotropy Probe (WMAP) three year results:
  %Implications for cosmology,''
  astro-ph/0603449.
  \bibitem{kklt}
  S.~Kachru, R.~Kallosh, A.~Linde and S.~P.~Trivedi,
  %``De Sitter vacua in string theory,''
  Phys.\ Rev.\ D {\bf 68}, 046005 (2003)
  [hep-th/0301240].
  %%CITATION = HEP-TH 0301240;%%

%\cite{landscape}
\bibitem{landscape}
  L.~Susskind,
  %``The anthropic landscape of string theory,''
  hep-th/0302219.
  \bibitem{Cohen:1998zx}
  A.~G.~Cohen, D.~B.~Kaplan and A.~E.~Nelson,
  %``Effective field theory, black holes, and the cosmological constant,''
  Phys.\ Rev.\ Lett.\  {\bf 82}, 4971 (1999)
  [hep-th/9803132].
  %%CITATION = HEP-TH 9803132;%%

%\cite{Horava:2000tb}
\bibitem{Horava:2000tb}
  P.~Horava and D.~Minic,
  %``Probable values of the cosmological constant in a holographic theory,''
  Phys.\ Rev.\ Lett.\  {\bf 85}, 1610 (2000)
  [hep-th/0001145];\\
  %%CITATION = HEP-TH 0001145;%%
%\cite{Thomas:2002pq}
%\bibitem{Thomas:2002pq}
  S.~D.~Thomas,
  %``Holography stabilizes the vacuum energy,''
  Phys.\ Rev.\ Lett.\  {\bf 89}, 081301 (2002).
  %%CITATION = PRLTA,89,081301;%%

%\cite{Hsu:2004ri}
\bibitem{Hsu:2004ri}
  S.~D.~H.~Hsu,
  %``Entropy bounds and dark energy,''
  Phys.\ Lett.\ B {\bf 594}, 13 (2004)
  [hep-th/0403052].
  %%CITATION = HEP-TH 0403052;%%

%\cite{Li:2004rb}
\bibitem{Li:2004rb}
  M.~Li,
  %``A model of holographic dark energy,''
  Phys.\ Lett.\ B {\bf 603}, 1 (2004)
  [hep-th/0403127].
  %%CITATION = HEP-TH 0403127;%%

%\cite{holoprin}
\bibitem{holoprin}
  G.~'t Hooft,
  %``Dimensional reduction in quantum gravity,''
  gr-qc/9310026;\\
  %%CITATION = GR-QC 9310026;%%
%\cite{Susskind:1994vu}
%\bibitem{Susskind:1994vu}
  L.~Susskind,
  %``The World as a hologram,''
  J.\ Math.\ Phys.\  {\bf 36}, 6377 (1995)
  [hep-th/9409089].
\bibitem{obs3}
  Q.~G.~Huang and Y.~G.~Gong,
  %``Supernova constraints on a holographic dark energy model,''
  JCAP {\bf 0408}, 006 (2004)
  [astro-ph/0403590];\\
  %%CITATION = ASTRO-PH 0403590;%%
%\bibitem{obs4}
  K.~Enqvist, S.~Hannestad and M.~S.~Sloth,
  %``Searching for a holographic connection between dark energy and the  low-l
  %CMB multipoles,''
  JCAP {\bf 0502} 004 (2005)
  [astro-ph/0409275];\\
  %%CITATION = ASTRO-PH 0409275;%%
%\cite{Shen:2004ck}
%\bibitem{Shen:2004ck}
  J.~Shen, B.~Wang, E.~Abdalla and R.~K.~Su,
  %``Constraints on the dark energy from the holographic connection to the
  %small l CMB suppression,''
  Phys.\ Lett.\ B {\bf 609} 200 (2005)
  [hep-th/0412227];\\
  %%CITATION = HEP-TH 0412227;%%
%\cite{Kao:2005xp}
%\bibitem{Kao:2005xp}
  H.~C.~Kao, W.~L.~Lee and F.~L.~Lin,
  %``CMB constraints on the holographic dark energy model,''
  Phys.\ Rev.\ D {\bf 71} 123518 (2005)
  [astro-ph/0501487].
  %%CITATION = ASTRO-PH 0501487;%%

%\cite{nonflat}
\bibitem{nonflat}
 Q.~G.~Huang and M.~Li,
 % ``The holographic dark energy in a non-flat universe,''
  JCAP {\bf 0408}, 013 (2004)
 [astro-ph/0404229].
  %%CITATION = ASTRO-PH 0404229;%%

% % Q.~G.~Huang and M.~Li,
  %``The holographic dark energy in a non-flat universe,''
 % JCAP {\bf 0408}, 013 (2004)
 % [astro-ph/0404229];\\
  %%CITATION = ASTRO-PH 0404229;%%
%\cite{Ito:2004qi}
%\bibitem{Ito:2004qi}
%  M.~Ito,
  %``Holographic dark energy model with non-minimal coupling,''
%  Europhys.\ Lett.\  {\bf 71}, 712 (2005)
%  [hep-th/0405281];\\
  %%CITATION = HEP-TH 0405281;%%
  %\cite{Enqvist:2004xv}
%\bibitem{Enqvist:2004xv}
  K.~Enqvist and M.~S.~Sloth,
  %``A CMB / dark energy cosmic duality,''
  Phys.\ Rev.\ Lett.\  {\bf 93}, 221302 (2004)
  [hep-th/0406019];\\
  %%CITATION = HEP-TH 0406019;%%
%\cite{Ke:2004nw}
%\bibitem{Ke:2004nw}
  K.~Ke and M.~Li,
  %``Cardy-Verlinde formula and holographic dark energy,''
  Phys.\ Lett.\ B {\bf 606}, 173 (2005)
  [hep-th/0407056];\\
  %%CITATION = HEP-TH 0407056;%%
%\cite{Huang:2004mx}
%\bibitem{Huang:2004mx}
  Q.~G.~Huang and M.~Li,
  %``Anthropic principle favors the holographic dark energy,''
  JCAP {\bf 0503}, 001 (2005)
  [hep-th/0410095];\\
  %%CITATION = HEP-TH 0410095;%%
%\cite{Zhang:2005yz}
%\bibitem{Zhang:2005yz}
%  X.~Zhang,
  %``Statefinder diagnostic for holographic dark energy model,''
 % Int.\ J.\ Mod.\ Phys.\ D {\bf 14}, 1597 (2005)
 % [astro-ph/0504586];\\
  %%CITATION = ASTRO-PH 0504586;%%
%\cite{Pavon:2005yx}
%\bibitem{Pavon:2005yx}
  D.~Pavon and W.~Zimdahl,
  %``Holographic dark energy and cosmic coincidence,''
  Phys.\ Lett.\ B {\bf 628}, 206 (2005)
  [gr-qc/0505020];\\
  %%CITATION = GR-QC 0505020;%%
%\cite{Wang:2005jx}
%\bibitem{Wang:2005jx}
  B.~Wang, Y.~Gong and E.~Abdalla,
  %``Transition of the dark energy equation of state in an interacting
  %holographic dark energy model,''
  Phys.\ Lett.\ B {\bf 624}, 141 (2005)
  [hep-th/0506069];\\
  %%CITATION = HEP-TH 0506069;%%
%\cite{Kim:2005at}
%\bibitem{Kim:2005at}
    S.~Nojiri and S.~D.~Odintsov,
  % ``Unifying phantom inflation with late-time acceleration: Scalar
  % phantom-non-phantom transition model and generalized holographic dark
  %energy,''
  Gen.\ Rel.\ Grav.\  {\bf 38}, 1285 (2006)
  [hep-th/0506212];\\
  %%CITATION = HEP-TH 0506212;%%
%\cite{Elizalde:2005ju}
%\bibitem{Elizalde:2005ju}
  E.~Elizalde, S.~Nojiri, S.~D.~Odintsov and P.~Wang,
   %``Dark energy: Vacuum fluctuations, the effective phantom phase, and
  %holography,''
  Phys.\ Rev.\ D {\bf 71}, 103504 (2005)
  [hep-th/0502082];\\
  %%CITATION = HEP-TH 0502082;%%
%\cite{Hu:2006ar}
%\bibitem{Hu:2006ar}
  B.~Hu and Y.~Ling,
  %``Interacting dark energy, holographic principle and coincidence problem,''
  Phys.\ Rev.\ D {\bf 73}, 123510 (2006)
  [hep-th/0601093];\\
  %%CITATION = HEP-TH 0601093;%%
%\cite{Li:2006ci}
%\bibitem{Li:2006ci}
  H.~Li, Z.~K.~Guo and Y.~Z.~Zhang,
  %``A tracker solution for a holographic dark energy model,''
  Int.\ J.\ Mod.\ Phys.\ D {\bf 15}, 869 (2006)
  [astro-ph/0602521];\\
    %%CITATION = HEP-TH 0609069;%%
%\cite{Setare:2006pj}
%\bibitem{Setare:2006pj}
  M.~R.~Setare,
  %``Bulk-brane interaction and holographic dark energy,''
  [hep-th/0609104];\\
M. R. Setare,  J. Zhang, X. Zhang, JCAP 0703,  007, (2007);\\
M. R. Setare, Phys. Lett. B {\bf648}, 329, (2007).
\bibitem{interac}
  L.~Amendola and D.~Tocchini-Valentini,  Phys.\ Rev.\ D {\bf 64}, 043509 (2001) [astro-ph/0011243];\\
  W.~Zimdahl, D.~J.~Schwarz, A.~B.~Balakin and D.~Pavon, Phys.\ Rev.\ D {\bf 64}, 063501 (2001) [astro-ph/0009353];\\
  A.~B.~Balakin, D.~Pavon, D.~J.~Schwarz and W.~Zimdahl, New J.\ Phys.\  {\bf 5}, 085 (2003) [astro-ph/0302150];\\
    R. Horvat,  Phys. Rev. D{\bf70}, 087301 (2004) [astro-ph/0404204];\\
    P.~Wang and X.~H.~Meng,    Class.\ Quant.\ Grav.\  {\bf 22}, 283 (2005) [astro-ph/0408495];\\
            B.~Wang, C.~Y.~Lin and E.~Abdalla, Phys.\ Lett.\  B {\bf 637}, 357 (2006) [hep-th/0509107];\\
    D.~Pavon and W.~Zimdahl,  AIP Conf.\ Proc.\  {\bf 841}, 356 (2006) [hep-th/0511053];\\
    M.~S.~Berger and H.~Shojaei, Phys.\ Rev.\ D {\bf 73}, 083528, (2006) [gr-qc/0601086].\\
    M.~R.~Setare,
  %``Interacting holographic dark energy model in non-flat universe,''
  Phys.\ Lett.\ B {\bf 642}, 1 (2006)
  [hep-th/0609069];\\
    M. R. Setare,  Eur. Phys. J. C {\bf50}, 991,  (2007);\\
    M. R. Setare, JCAP, 0701, 023 (2007);\\
M. R. Setare, 0708.0118 [hep-th], accepted for publication in Phys.
Lett. {\bf B}(2007).
\bibitem{gong}Y. Gong,  Phys. Rev. D, {\bf70}, 064029, (2004).
\bibitem{mu} H. Kim, H. W. Lee, and Y. S. Myung,
Phys. Lett. B {\bf628}, 11, (2005).
\bibitem{tor} D. F. Torres, Phys. Rev. {\bf D66}, 043522, (2002).
\bibitem{set1}M. R. Setare, Phys. Lett. B {\bf 644}, 99, (2007).
\bibitem{bd}C. Brans and C. H. Dicke, Phys. Rev. {\bf124}, 925 (1961).
\bibitem{1}S. Capozziello, Int. J. Mod. Phys. D 11, 483 (2002); S. Capozziello,
 S. Carloni and A. Troisi, arXiv:astro-ph/0303041;
S. M. Carroll, V. Duvvuri, M. Trodden and S. Turner, Phys. Rev. D 70
(2004) 043528.
\bibitem{2}S. Nojiri and S. D. Odintsov, Phys. Rev. D 68, 123512 (2003) [arXiv:hep-th/0307288].
\bibitem{3}S. Nojiri, S. D. Odintsov and M. Sasaki, Phys. Rev. D 71, 123509 (2005)[arXiv:hep-th/0504052].
\bibitem{Kim:2005at}H.~Kim, H.~W.~Lee and Y.~S.~Myung,
    Phys.\ Lett.\ B {\bf 632}, 605 (2006)
  [gr-qc/0509040].
\bibitem{HG}  Q. G. Huang, Y. Gong, JCAP, 0408, (2004),006.
\bibitem{cmb1} H. C. Kao, W. L. Lee and F. L. Lin,
 astro-ph/0501487.
\bibitem{cmb3}  J. Shen, B. Wang, E. Abdalla and R. K. Su,
 hep-th/0412227.
\end{thebibliography}
\end{document}